\documentclass[osajnl,twocolumn,showpacs]{revtex4}

\usepackage{graphicx}
\graphicspath{{pictCoupler/}{}}

\newcommand\pictc[5]{\begin{figure}
            \centerline{\vspace{0mm}
\includegraphics[width=\columnwidth,height=0.7\textheight,keepaspectratio]{#3}}
            \protect\caption{\protect\label{fig:#4} #5}
                    \end{figure}            }

\newcommand\pict[4][1]{\pictc{#1}{!tb}{#2}{#3}{#4}}
\newcommand\rpict[1]{\ref{fig:#1}}

\newcommand\leqt[1]{\protect\label{eq:#1}}
\newcommand\reqtn[1]{\ref{eq:#1}}
\newcommand\reqt[1]{(\reqtn{#1})}

\newcounter{Fig}

\begin{document}
\begin{sloppy}

\title{Nonlinear directional coupler for polychromatic light}

\author{Ivan L. Garanovich}
\author{Andrey A. Sukhorukov}

\affiliation{Nonlinear Physics Centre and Centre for Ultra-high bandwidth Devices for Optical Systems (CUDOS),\\
 Research School of Physical Sciences and Engineering, Australian National University, Canberra, ACT 0200, Australia}

\begin{abstract}
We demonstrate that nonlinear directional coupler with special bending of waveguide axes can be used for all-optical switching of polychromatic light with very broad spectrum covering all visible region. The bandwidth of suggested device is enhanced five times compared to conventional couplers. Our results suggest novel opportunities for creation of all-optical logical gates and switches for polychromatic light with white-light and super-continuum spectrum.
\end{abstract}


\maketitle

Directional waveguide coupler
first introduced by Jensen~\cite{Jensen:1982-1568:ITMT} and Maier~\cite{Maier:1982-2296:KE} has attracted a great deal of attention
as a major candidate for creation of ultra-fast all-optical switches.
This device utilizes light tunneling between two optical waveguides placed in close proximity to each other, as schematically shown in Fig.~\rpict{Coupling}(a). In the linear regime, light is switched from one to another waveguide at the distance called coupling length. At high input powers, intensity-dependent change of the refractive index through optical nonlinearity creates detuning between the waveguides which can suppress power transfer between coupler arms, such that light remains in the input waveguide.
Since the first experimental demonstration of a subpicosecond nonlinear coupler switch in a dual-core fiber~\cite{Friberg:1987-1135:APL}, various aspects of switching in different coupler configurations has been  extensively analyzed~\cite{Leutheuser:1990-251:OC, Assanto:1993-1323:APL, Karpierz:1995-61:PAO, Skinner:1995-493:OC, Betlej:2006-1480:OL}.

In recent years, new sources of light with ultra-broad spectrum became available, having a wide range of applications including information transmission, spectroscopy, microscopy, and optical sensing. However, conventional coupler can only perform switching of signals with rather limited spectral bandwidth, because the coupling length depends on optical frequency resulting in separation of different frequency components between the waveguides.

In this Letter, we propose a new configuration of directional coupler designed for nonlinear switching of polychromatic light, such as light with supercontinuum frequency spectrum generated in photonic-crystal fibers and fiber tapers~\cite{Ranka:2000-25:OL, Wadsworth:2002-2148:JOSB}. The spectral bandwidth of suggested device is five times wider compared to conventional coupler structures, making it possible to collectively switch wavelengths covering almost all visible region.

\pict{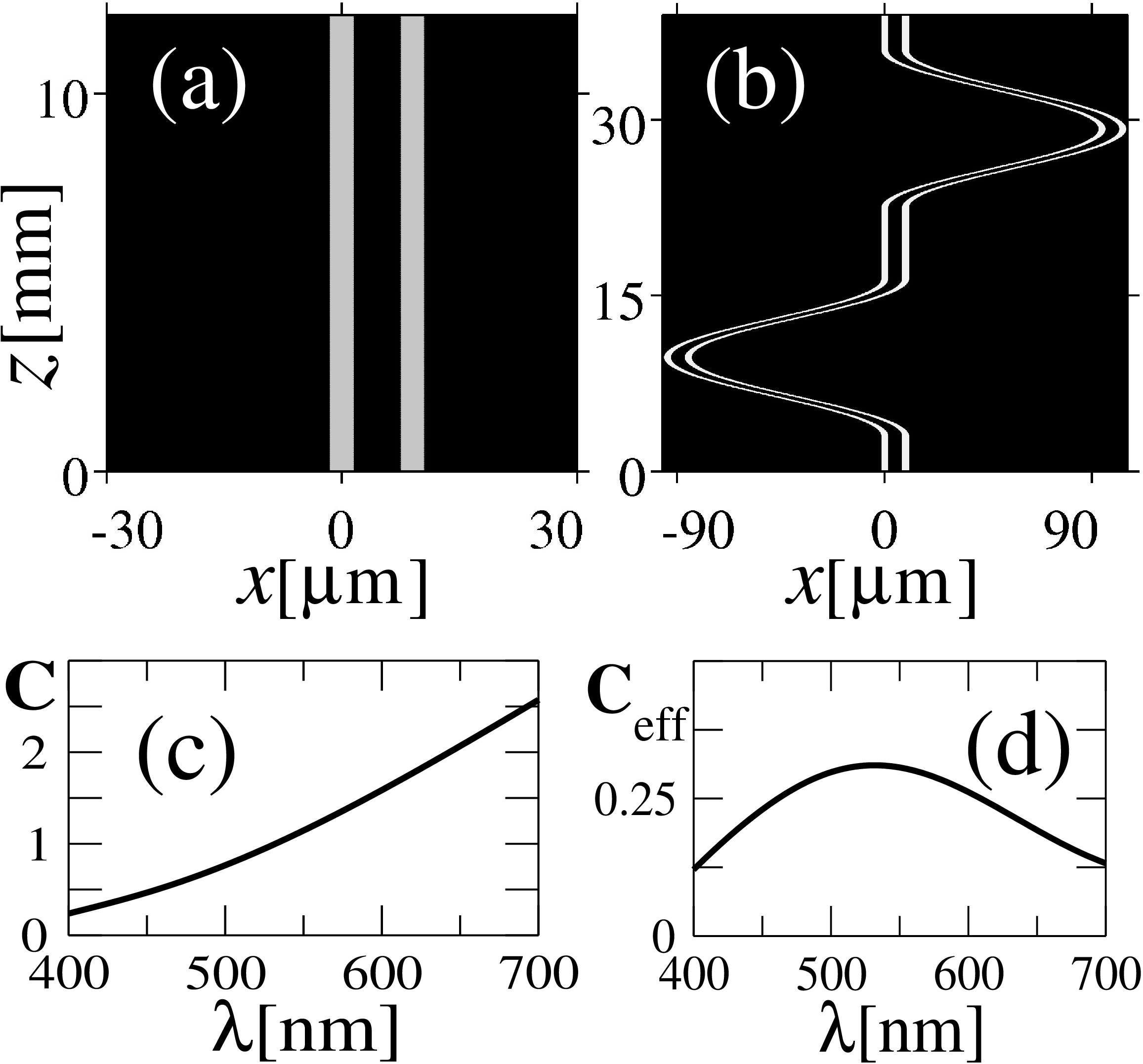}{Coupling}{
(a)~Conventional directional coupler composed of two evanescently coupled straight waveguides.
(b)~Polychromatic coupler with specially designed bending of the waveguide axes.
(c)~Wavelength-dependence of the coupling coefficient between straight waveguides.
(d)~Effective coupling in the curved coupler shown in~(b).
Waveguide width and separation between waveguide axes are $3\mu$m and $9\mu$m, respectively. Refractive index contrast is $\Delta \nu = 8 \times 10^{-4}$, and $n_0 =2.35$.
}

We demonstrate that the operating bandwidth of conventional coupler consisting of straight parallel waveguides [Fig.~\rpict{Coupling}(a)] can be improved by introducing special bending of waveguide axes in the propagation direction as illustrated in Fig.~\rpict{Coupling}(b).
Nonlinear switching of polychromatic signals while preserving their spectral characteristics can be realized in media with slow nonlinear response, where the optically-induced refractive index change is defined by the time-averaged light intensity of different spectral components~\cite{Mitchell:1997-880:NAT, Buljan:2004-397:JOSB}. Then, the evolution of polychromatic beams can be described by a set of normalized nonlinear equations for the spatial beam envelopes $A_m(x,z)$
at vacuum wavelengths $\lambda_m$,
\begin{equation} \leqt{nls}
   i \frac{\partial A_m}{\partial z}
   + \frac{z_s \lambda_m}{4 \pi n_0 x_s^2} \frac{\partial^2 A_m}{\partial x^2}
   + \frac{2 \pi z_s}{\lambda_m}
          \left\{ \nu\left[x-x_0(z)\right] + {\cal G} \right\} A_m = 0 ,
\end{equation}
where $x$ and $z$ are the transverse and propagation coordinates normalized to the characteristic values $x_s = 1 \mu$m and $z_s = 1 mm$, respectively,
$n_0$ is the average refractive index of the medium,
$\nu(x)$ defines the transverse refractive index profile in cross-section of the coupler,
$x_0(z)$ is the longitudinal bending profile of the waveguide axes,
${\cal G} = \alpha M^{-1} \sum_{m=1}^M \gamma(\lambda_m) |A_m|^2$ defines nonlinear change of refractive index, $\alpha$ is the nonlinear coefficient, and $\gamma(\lambda)$ accounts for dispersion of the nonlinear response. In numerical simulations, we choose a large number of components $M = 50$ to model accurately the dynamics of beams with broadband spectrum.

\pict{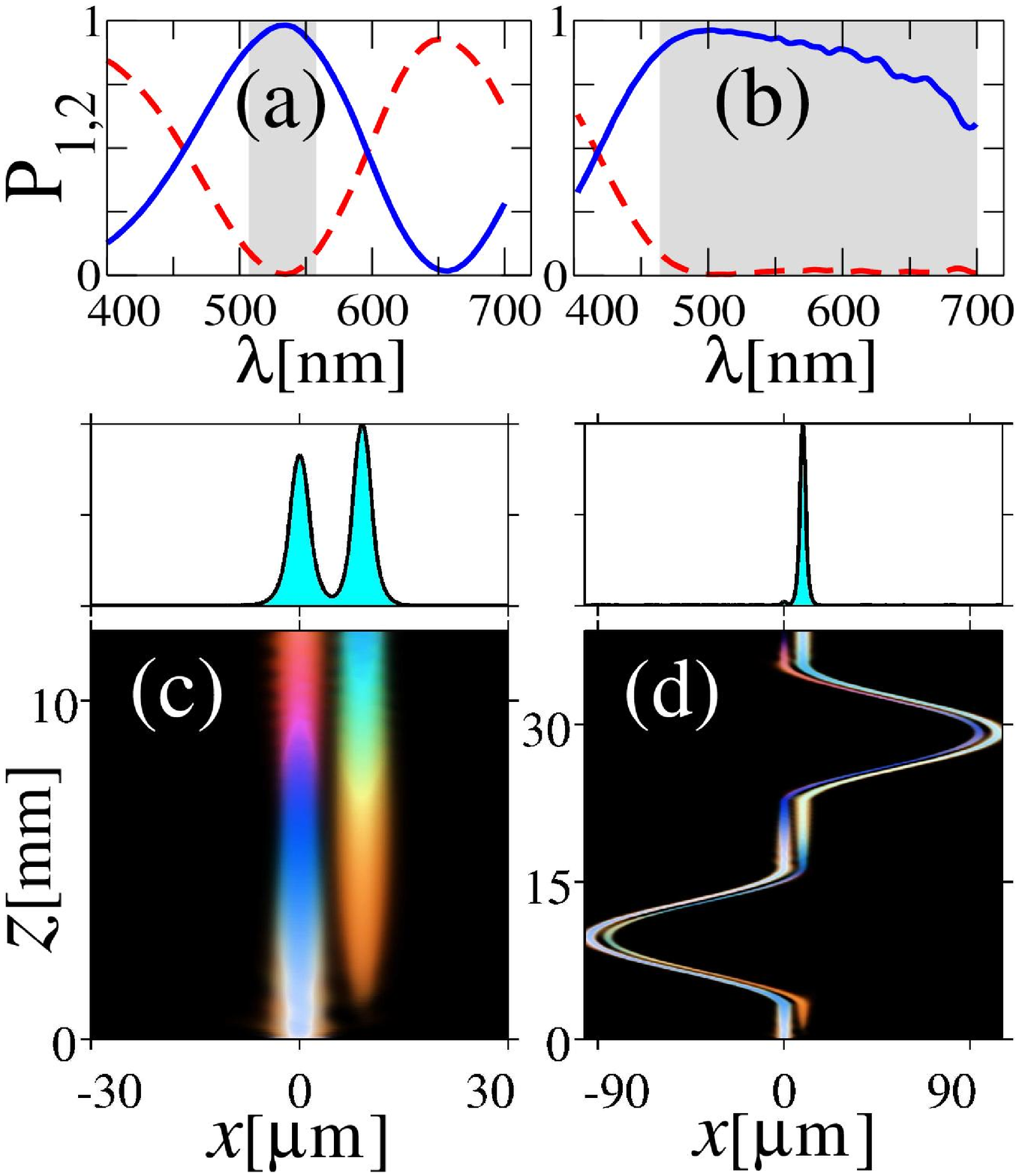}{LinearSwitching}{
(Color online)
(a,b)~Wavelength dependence of linear transmission characteristics for straight and optimized curved couplers, respectively. Shown are output powers in the left (dashed curve, $P_1$) and right (solid curve, $P_2$) coupler arms, when light is input at the left arm. Shading marks spectral regions where the switching ratio $P_2 / P_1$ is larger than 10.
(c,d)~Evolution of polychromatic light with flat spectrum covering 450-700 nm
in the straight and in the optimized curved structures, respectively. Top panels in~(c) and~(d) show the total intensity distributions at the output.
}

As monochromatic light propagates in a directional coupler made of straight identical waveguides [Fig.~\rpict{Coupling}(a)], the power is periodically exchanged between the two waveguides~\cite{Jensen:1982-1568:ITMT}. The period is defined by the coupling length, $Z_c = \pi / [2 C(\lambda)]$, where $C(\lambda)$ is the coupling coefficient. Then, signal switching between output coupler arms is realized by choosing the device length as an odd number of coupling lengths. However, this condition cannot be simultaneously satisfied for all frequency components of polychromatic light, since the coupling coefficient depends on wavelength~\cite{Mendes:2004-425:OC, Jensen:1982-1568:ITMT} and tends to increase at the red spectral edge, see Fig.~\rpict{Coupling}(c).

The effect of axes bending on light propagation in two coupled waveguides can be described in terms of the effective coupling coefficient $C_{\rm eff}$. It was shown~\cite{Longhi:2005-65801:PRA} that, in the limit when bending period is
smaller than the coupling length for straight waveguides, the light distribution at the output of the curved coupler is the same as for straight structure with the coupling $C_{\rm eff}$ between the waveguides.
For bending profiles consisting of symmetric segments with ${x}_0(z - \tilde{z}) = {x}_0(\tilde{z} - z)$, where $\tilde{z}$ is a coordinate shift, the modified coupling coefficient takes the same form as in periodic waveguide arrays~\cite{Longhi:2005-65801:PRA, Longhi:2006-243901:PRL}
$C_{\rm eff}( \lambda )
   = C( \lambda ) L^{-1}
      \int_{0}^{L} \cos\left[ 2\pi n_0 a \dot{x}_0(z) / \lambda \right] dz$.
Here $C(\lambda)$ is the wavelength-dependent coupling coefficient between straight waveguides with the same separation $a$ between their axes,
and the dot stand for the derivative.

We note that the integral in the expression for the effective coupling coefficient depends on the wavelength $\lambda$. This makes it possible to compensate for the wavelength-dependence of the coupling coefficient $C(\lambda)$ with the geometrical bending-induced dispersion. In particular, we can obtain almost the same effective coupling in a broad spectral range around the central wavelength $\lambda_0$ by satisfying the condition  $\left. d C_{\rm eff}(\lambda) / d \lambda \right|_{\lambda = \lambda_0} = 0$. We find that such wavelength-insensitive effective coupling can be realized in a hybrid coupler structure consisting of several alternating straight and sinusoidal sections,
$x_0(z) =  0$ for $0 \leq z \leq z_0$,
$x_0(z) =  A \lbrace \cos\left[2\pi (z - z_0) / (L/N - 2 z_0)\right] - 1 \rbrace$ for $z_0 \leq z \leq (L/N - z_0)$,
$x_0(z) =  0$ for $(L/N - z_0) \leq z \leq L/N$,
and $x_0(z) = - x_0(z-L/N)$ for $L/N \leq z \leq L$, where $N$ is the number of segments. An example of two-segment ($N=2$) structure is shown in Fig.~\rpict{Coupling}(b).
We set $A = \xi_2 (L/N - 2 z_0) \lambda_0 (4 \pi^2 n_0 a)^{-1}$,
where $\xi_2 \simeq 5.52$ is the second root of the equation $J_0(\tilde{\xi})=0$ and $J_m$ is the Bessel function of the first kind of the order $m$.
Effective coupling in this structure can be calculated analytically,
$C_{\rm eff}(\lambda) = C(\lambda) [ 2 N z_0 L^{-1}  + (1 - 2 N z_0 L^{-1} ) J_0(\xi_2 \lambda_0/\lambda) ]$, and condition $\left. d C_{\rm eff}(\lambda) / d \lambda \right|_{\lambda = \lambda_0} = 0$ is satisfied for $2 N z_0 L^{-1} = \left[ 1 - (\xi_2 J_1(\xi_2) C_0)^{-1} C_1 \right]^{-1}$. Note that the effective coupling does not depend on $N$. Here the coefficients $C_0 = C(\lambda_0)$ and
$C_1 = \lambda_0 \left. d C(\lambda) / d \lambda\right|_{\lambda = \lambda_0}$
characterize dispersion of coupling between straight waveguides.
In our numerical simulations, we choose the central wavelength at $\lambda_0 = 532$nm, and find the coupling dispersion for waveguides shown in Fig.~\rpict{Coupling}(a) as $C_0 \simeq 0.13$mm$^{-1}$ and $C_1 \simeq 0.52$mm$^{-1}$. Then, we calculate the optimal parameters of the curved coupler,
and obtain almost constant coupling $C_{\rm eff}(\lambda \simeq \lambda_0) \simeq  0.31 C_0$ in a broad spectral region, see Fig.~\rpict{Coupling}(d).

The optimized curved coupler can be used to collectively switch all spectral components around the central wavelength $\lambda_0$ from one input waveguide to the other waveguide at the output. This regime is realized when the device length is matched to the effective coupling length, i.e. $L = \pi / \left[ 2 C_{\rm eff}(\lambda_0) \right] \simeq 39$mm. The accuracy of the effective coupling approximation increases for larger $N$, and we found that the deviation from exact solutions of coupled-mode equations~\cite{Longhi:2005-65801:PRA} in the vicinity of central wavelength is less than 0.5\% for $N\ge 2$. Since bending losses decrease for larger $L/N$, we choose a two-segment ($N=2$) structure configuration [see Fig.~\rpict{Coupling}(b)]. We then perform numerical simulations based on full model Eq.~\reqt{nls} and confirm that the proposed coupler structure indeed exhibits extremely efficient switching into the crossed state simultaneously in a very broad spectral region of about $450-700$nm, which covers almost all visible, see Figs.~\rpict{LinearSwitching}(b) and (d). This is in a sharp contrast to the conventional straight coupler [Figs.~\rpict{LinearSwitching}(a) and (c)] that can only operate in the spectral region of $\sim 510-560$nm, which is about five time less than for the proposed curved coupler. We note that slight decrease of the output power at the red edge of the spectrum for the curved coupler [Fig.~\rpict{LinearSwitching}(b)] is caused by the radiation at the waveguide bends~\cite{Longhi:2005-65801:PRA}, but such losses do not affect the broadband switching behavior.

\pict{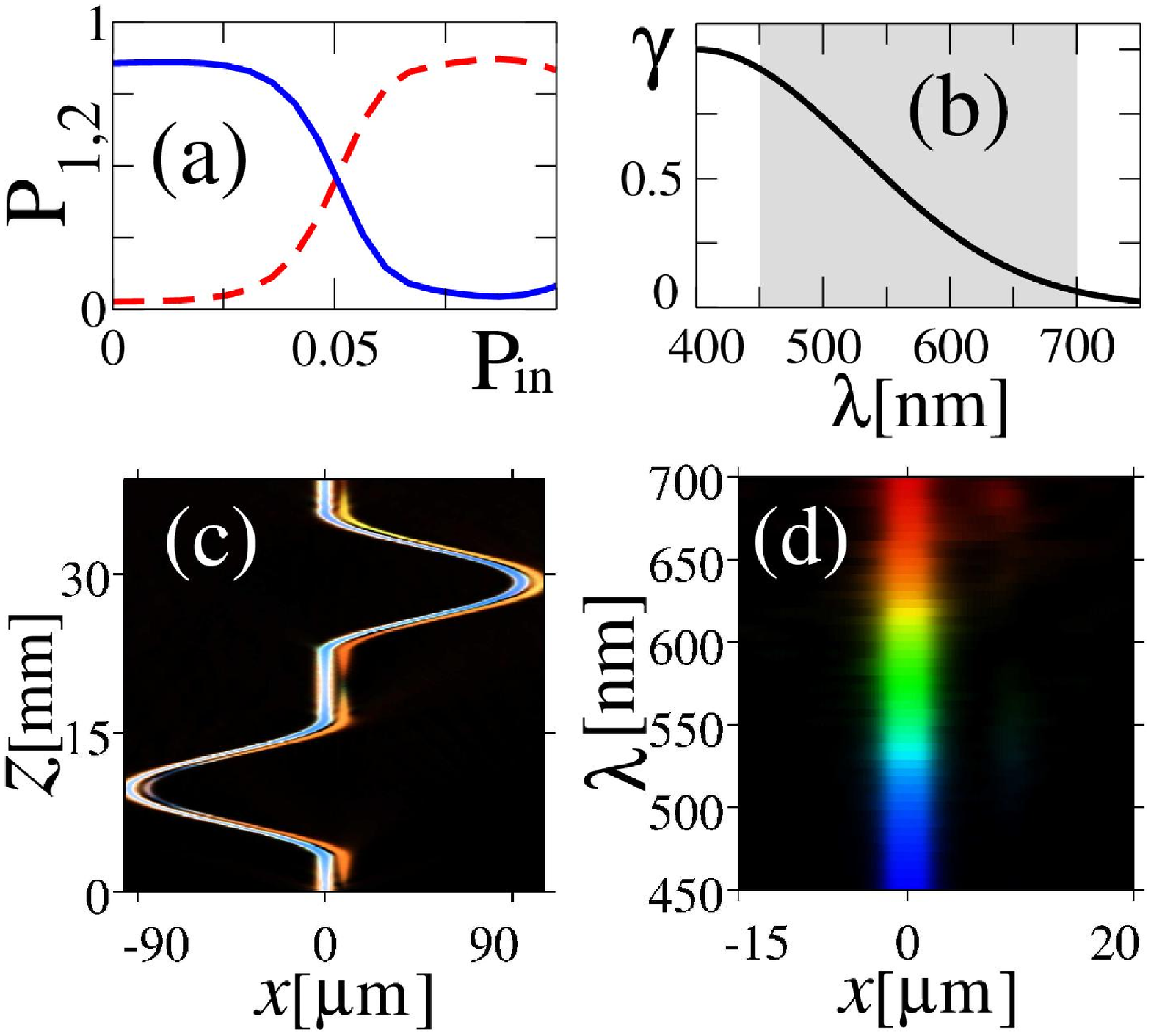}{NonlinearSwitching}{
(Color online) Nonlinear switching of polychromatic light.
(a)~Power distribution at the output ports of the coupler as a function of the input power. Polychromatic input is the same as in Figs.~\rpict{LinearSwitching}(c) and~ (d). Solid and dashed curves show power in the left ($P_1$) and in the right ($P_2$) output coupler ports, respectively.
(b)~Sensitivity function $\gamma$ describing wavelength-dispersion of the nonlinear response.
(c,d)~Propagation dynamics and output spectrum, respectively, in the nonlinear switched state realized at the total input power $P_{in} = 0.085$. Nonlinear coefficient is $\alpha = 10$.
}

At high input powers, nonlinear change of the refractive index modifies waveguide propagation constant and decouples waveguides from each other similar to other nonlinear coupler structures studied before~\cite{Jensen:1982-1568:ITMT, Mendes:2004-425:OC, Friberg:1987-1135:APL}. This causes switching from crossed state into the parallel state as shown in Figs.~\rpict{NonlinearSwitching}(a), (c) and (d). Remarkably, nonlinear switching also takes place in a very broad spectral region $\sim 450-700$nm, which enables the coupler to act as an all-optical digital switch for polychromatic light. In these simulations, we consider the case of a photorefractive medium such as LiNbO$_3$ where optical waveguides of arbitrary configuration can be fabricated by titanium indiffusion~\cite{Chen:2005-4314:OE, Matuszewski:2006-254:OE}. The photosensitivity of LiNbO$_3$ can be approximated as a function $\gamma(\lambda) = \exp[ -{\rm log}(2) (\lambda - \lambda_b)^2 / \lambda_w^2]$ which has a maximum at the wavelength $\lambda_b = 400nm$ and then drops two times with the $\lambda_w = 150nm$ shift to the red edge of the spectrum as shown in Fig.~\rpict{NonlinearSwitching}(b). We have verified that switching behavior of the coupler remains essentially the same for different values of $\lambda_w$, which primarily affect the quantitative characteristics of the coupler such as the switching power.

In conclusion, we demonstrated that optimized curved directional coupler can be used to perform switching of polychromatic light with extremely broad spectrum covering almost all visible. Similar principles can be applied to create broadband switches for other spectral regions. Suggested devices can be fabricated in planar waveguiding structures, offering novel opportunities for creation of all-optical logical gates and switches for polychromatic signals with white-light or super-continuum spectrum.

Authors thank Yuri Kivshar for useful discussions and comments. This work has been supported by the Australian Research Council. I. Garanovich's e-mail address is ilg124@rsphysse.anu.edu.au.

\end{sloppy}
\end{document}